\documentclass[aps,pra,reprint,twocolumn,superscriptaddress,amsmath,amssymb,citeautoscript]{revtex4-1}

\usepackage{amsmath,amssymb}
\usepackage[usenames]{color}
\usepackage{amssymb}
\usepackage{grffile}
\usepackage[pdftex]{graphicx}
\usepackage{amsmath, amstext, amssymb, amsfonts, amsxtra}
\usepackage{textcomp}
\usepackage{xspace}
\usepackage{color}

\usepackage[colorlinks]{hyperref}

\newcommand\e{{\rm e}}
\newcommand\ii{{\rm i}}
\newcommand\HHop{\mathcal{H}}

\newcommand{\hannover}{Institut f\"ur Theoretische Physik, Leibniz Universit\"at Hannover, Appelstr. 2, DE-30167 Hannover, Germany}

\begin{document}

\title{Probing the exchange statistics of one-dimensional anyon models}


\author{Sebastian Greschner} 
\affiliation{Department of Quantum Matter Physics, University of Geneva, CH-1211 Geneva, Switzerland} 
\author{Lorenzo Cardarelli} 
\affiliation{\hannover}
\author{Luis Santos}
\affiliation{\hannover}  

\date{\today}

\begin{abstract}
We propose feasible scenarios for revealing the modified exchange statistics in one-dimensional anyon models in optical lattices based 
on an extension of the multicolor lattice-depth modulation scheme introduced in [{Phys. Rev. A 94, 023615 (2016)}].
We show that the fast modulation of a two-component fermionic lattice gas in the presence a magnetic field gradient, in combination with additional resonant microwave fields, 
allows for the quantum simulation of hardcore anyon models with periodic boundary conditions. Such a semi-synthetic ring set-up allows for realizing an interferometric arrangement 
sensitive to the anyonic statistics. Moreover, we show as well that simple expansion experiments may reveal the formation of anomalously bound pairs resulting from the anyonic 
exchange.
\end{abstract}

\maketitle



\section{Introduction}
\label{sec:introduction}

Anyons~\cite{Wilczek1982,Leinaas1977}, particles with an exchange statistics interpolating between bosons and fermions, play a crucial role in fascinating concepts of modern condensed matter physics, such as topological quantum phases, in particular fractional quantum Hall effect~\cite{Laughlin1983,Halperin1984,Camino2005,Kim2005} and topological quantum computing~\cite{Kitaev2003, Nayak2008}. 
The experimental search and manipulation of anyons has attracted a large interest in recent years, including spin and boson models~\cite{Kitaev2003, Shen2014,Pan2003, Zhang2008, Feng2013}, and systems of ultracold atoms~\cite{Paredes2001,Duan2003,Micheli2006,Aguado2008,Jiang2008}. While theoretically settled, an unambiguous detection of anyonic (quasi-) particles, e.g. by interferometric measurements~\cite{Camino2005}, 
is still the object of active research~\cite{Jiang2008, Bonderson2008}.

Although anyons were originally proposed for two dimensional~(2D) systems, one-dimensional~(1D) anyons have been theoretically studied as well~\cite{Haldane1991, Ha1994, Murthy1994, Wu1995, Zhu1996, Amico1998,Kundu1999, Batchelor2006, Girardeau2006}. The exotic properties of 1D (Abelian) anyon models include asymmetric momentum distributions~\cite{Calabrese2007,Patu2007,Hao2008,Hao2009,Calabrese2009,Tang2015}, particle dynamics~\cite{DelCampo2008,Hao2012,Wang2014}, entanglement properties~\cite{Santachiara2007,Guo2009, Marmorini2016} or statistically induced Mott insulator to superfluid quantum phase transitions~\cite{Keilmann2011,Greschner2015,Forero2016,Zhang2017}.
Despite this theoretical interest, the experimental realization of 1D anyons is still missing. 
A (pseudo) anyon Hubbard model (AHM) in 1D optical lattices may be engineered by means of Raman-assisted tunneling~\cite{Keilmann2011, Greschner2014AB,Greschner2015anyon}. Pseudo anyons exhibit anyonic commutation off-site, but on-site behave as bosons, i.e. there may be more than one particle per lattice site. A drastically simplified realization of the AHM may be realized 
by means of periodically driven lattices~\cite{Straeter2016}. A proper three-color modulation of the lattice depth has been proposed for the realization of a two-component 1D anyon Hubbard model (2AHM)~\cite{Cardarelli2016}.

As for 2D anyons, revealing the  anyonic character of the engineered 1D quasi-particles remains an interesting open question.
This paper proposes two feasible experiments in which the modified statistics may be revealed in ultra-cold lattice gases.
On the one hand, we show that expansion experiments using the three-color modulation of Ref.~\cite{Cardarelli2016} may reveal the formation of anomalous bound-state pairs. These pairs, 
which result from the anyonic exchange statistics, anticipate the emergence of the exotic partially paired phase~(PP) predicted for the 2AHM~\cite{Cardarelli2016}.
On the other hand, combining the three-color modulation with spin-dependent tilting and Raman-assisted coupling at the system boundaries allows 
for the realization of an effective one-component hardcore (i.e. with at most one particle per site) AHM with periodic boundary conditions~(PBC). This effective synthetic ring
set-up allows for interferometric measurements that reveal the statistical angle that characterizes the anyonic character of the particles.

The paper is organized as follows. After an introduction to the three-color modulation scheme in Sec.~\ref{sec:3color} and a discussion of the mappings to anyon models in the low-density limit, Sec.~\ref{sec:interferometer} is devoted to the expansion of particles in the hardcore AHM with PBC, and its characteristic dependence on the statistical angle. Finally in Sec.~\ref{sec:2AHMexpand} we discuss expansion 
experiments for the 2AHM model, and conclude in Sec.~\ref{sec:outlook} with a summary and a short outlook.

\section{Three-color modulation of an interacting Fermi-gas}
\label{sec:3color}

\begin{figure}[tb]
\begin{center}
\includegraphics[width=1\linewidth]{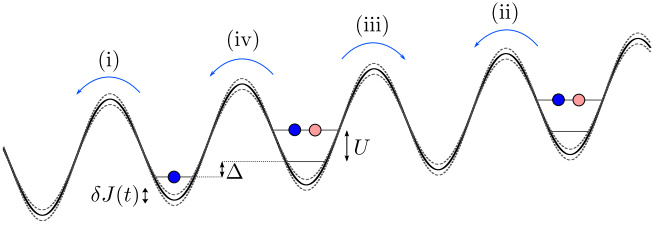}
\caption{Sketch of the lattice set-up with a spin independent tilting $\Delta$ and the relevant hopping processes (i)-(iv). Blue (red) bullets correspond to spin $\uparrow$ ($\downarrow$) particles.}
\label{fig:scheme1}
\end{center}
\end{figure}

\begin{figure}[tb]
\begin{center}
\includegraphics[width=1\linewidth]{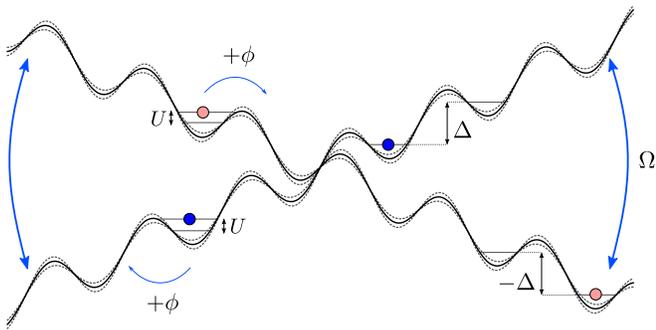}
\caption{Lattice shaking scheme with a magnetic field gradient for the realization of the hardore anyon hubbard model. Microwave fields $\Omega$ couple the boundaries of the system.}
\label{fig:scheme2}
\end{center}
\end{figure}

In the following we will recapitulate and extend the multicolor modulation scheme introduced in Ref.~\cite{Cardarelli2016}. The main experimental idea is depicted in Figs.~\ref{fig:scheme1} and \ref{fig:scheme2}. A two-component ($\sigma=0,1$ corresponding to spin $\uparrow$ and $\downarrow$) Fermi (or hardcore Bose) gas is loaded into a 1D optical lattice. 
We assume a tilted lattice, with an energy shift $\Delta$ between neighboring sites~(Fig.~\ref{fig:scheme1}) due to acceleration or tilting in gravity, and, hence, the tilting is spin independent. An interesting alternative 
is to employ a magnetic field gradient, which leads to a spin dependent tilting~(Fig.~\ref{fig:scheme2}).

The system is then described by the Fermi-Hubbard Hamiltonian:
\begin{equation}
\HHop(t)\!=\! -J(t)\sum_{j,\sigma} \! \left [ c_{j+1,\sigma}^\dag c_{j,\sigma}\! +\! \mathrm{H.c.} \right ]\! + \HHop_{\rm int} + \HHop_{\rm tilt},
\label{eq:H_full_J(t)}
\end{equation}
where $c_{j,\sigma}$ is the annihilation operator of a fermion with spin  $\sigma$ at site $j$, and the tilting is given by
\begin{align}
\HHop_{\rm tilt} \!=\! \Delta \sum_{j,\sigma} \epsilon_\sigma j n_{j,\sigma}\,,
\end{align}
where $\epsilon_\sigma=1$ for the spin-independent tilting and $(-1)^\sigma$ for the spin-dependent case.
Importantly, we assume that the particles experience a repulsive on-site interaction, 
\begin{align}
\HHop_{\rm int} \!=\! U \sum_j n_{j,\uparrow}n_{j,\downarrow} \,.
\end{align}

In both cases we assume that a direct hopping of the particles between the lattice sites can be neglected, $J(t) \ll \Delta, |\Delta\pm U| $.
The hopping is restored by a resonant modulation the optical lattice. Following Ref.~\cite{Cardarelli2016} we assume a fast periodic modulation of the optical lattice depths~\cite{Ma2011}, $V(t)=V_0+\delta V(t)$, with $\delta V\ll V_0$.
One could equivalently assume a fast lattice shaking such as discussed in Ref.~\cite{Straeter2016}. 
 We may then integrate out the fast periodic drivings and obtain the effective anyon models as we will discuss in the following.



\subsection{The spin-independent tilting}

From Fig.~\ref{fig:scheme1} we may identify four relevant hopping processes in the spin-independent tilting which we would like to restore by means of the resonant drivings and the corresponding energy differences $\Delta E_i$:
\begin{itemize}
\item (i) a single atom hops to an empty site to its right $\Delta E_1\!=\!\Delta$; 
\item (ii) a single atom tunnels to occupied site at its right: $\Delta E_{2}\!=\!\Delta\!+\!U$; 
\item (iii) same as (ii) but the hopping is to the left: $\Delta E_{3}\!=\! U\! -\!\Delta$; 
\item  (iv) an atom in a doubly-occupied site tunnels into a single-occupied site~("doublon hopping"): $\Delta E_{4}\!=\!\Delta$
\end{itemize}

As discussed in Ref.~\cite{Cardarelli2016}, a separate but simultaneous driving with three frequencies,  $\omega_1=\Delta$, $\omega_2=\Delta+U-\tilde U$, and $\omega_3=-\Delta+U-\tilde U$, allows for (quasi-) resonantly restoring the four hopping processes. The detuning $|\tilde U|\ll U$ allows for the introduction of an effective two-body interaction. In the following we will set $\tilde{U}=0$.
Hence, we choose a modulation of the laser intensity as 
\begin{align}
\delta V(t)=\delta V \sum_{s=1,2,3} \cos(\omega_s t+\phi_s)
\end{align}
which corresponds to a modulation of the tunneling amplitude as
\begin{align}
\delta J(t)= \delta J \sum_s \cos(\omega_s t+\phi_s)\,.
\end{align}
An important aspect is here, that we may arbitrarily choose the phases $\phi_s$ for all three processes, which can be exploited to realize the fractional statistics.

After integrating out the resonant driving we recover a model without tilting
\begin{align}
\!\HHop_{\text{eff}}\!=\! -\frac{\delta J}{2} \!\sum_{j,\sigma}\! c_{j+1,\sigma}^\dag \e^{i \phi |n_{\j+1,\bar\sigma}\!-\! n_{j,\bar\sigma}|} c_{j,\sigma} + \dots,
\label{eq:H_eff}
\end{align}
Interestingly, within the scheme we may also control the amplitude of the three drivings separately which opens further possibilities, e.g. allows for the simulation of more general correlated hopping Hubbard models with asymmetric hopping amplitudes for doublons and single particles. Here, however we focus on the properties of the AHM with symmetric hoppings.
As discussed in Ref.~\cite{Cardarelli2016} higher order terms in this scenario (the ellipsis in Eq.~\eqref{eq:H_eff}), may induce effective interactions between nearest neighbor~(NN) sites due to the virtual hopping of particles. Those NN interactions open additional interesting possibilities for the observation of interacting quantum gases. While this aspect was analyzed in detail in Ref.~\cite{Cardarelli2016}, here in the following we will neglect this issue.

At low lattice filling $\rho$, for which processes (iv) may be neglected,  
a Jordan-Wigner like transformation~\cite{Keilmann2011},
\begin{equation}
f_{j,\sigma}=\e^{-\ii 2 \phi \sum_{1\leq l< j} n_l} \e^{-\ii \phi n_j} c_{j,\sigma},
\end{equation}
maps model Eq.~\eqref{eq:H_eff} into a 2AHM:
\begin{eqnarray}
\!\!\!\HHop_{\text{2AHM}}\! \! &=&\! \!  - \frac{\delta J_1}{2} \sum_{j,\sigma} (f_{j,\sigma}^{\dagger}f_{j+1,\sigma}^{\phantom \dagger} \!+\! \text{H.c.}) \!+\! \tilde{U}\HHop_{\text{int}}.
\label{eq:2AHM}
\end{eqnarray}
The operators $f_{j,\sigma}$ and $f_{j,\sigma}^\dagger$ characterize anyon-like hardcore particles, that fulfill a deformed exchange statistics:
\begin{align}
&f_{j,\sigma} f_{k,\sigma'}^\dagger + \mathcal{F}_{j,k} f_{k,\sigma'}^\dagger f_{j,\sigma} = \delta_{j,k} \delta_{\sigma,\sigma'} , \nonumber\\
&f_{j,\sigma} f_{k,\sigma'} + \mathcal{F}_{j,k} f_{k,\sigma'} f_{j,\sigma} = 0 .
\end{align}
The  complex parameter $\mathcal{F}_{j,k}$ determines the statistics of the system:
\begin{eqnarray}
&{\cal F}_{j,k}:=\left \{ \begin{array}{ll} e^{-\ii 2 \phi},\ \quad & j>k,  \\
1, \quad & j=k, \\
e^{\ii 2 \phi}, \quad & j<k,  \end{array} \right.  \quad &
\end{eqnarray}
where the condition ${\cal F}_{j,j}=1$ sets the hard-core behavior of the particles.
Note that for $\phi=0$ we retrieve the two-component Fermi-Hubbard model, while $\phi=\pi/2$ corresponds to the two-component hard-core Bose-Hubbard model.
Non-trivial quantum effects may be observed even for $\tilde{U}=0$ and $\phi=\pi/2$~\cite{Cardarelli2016}.

\subsection{Magnetic field gradient}

We now consider the case of a spin dependent tilting of the optical lattice such as realized in a magnetic field gradient Fig.~\ref{fig:scheme2}. Again we may identify four hopping processes and three corresponding frequencies.
\begin{itemize}
\item (i) single atom hops to an empty site to its right $\Delta E_1\!=\!\Delta$; 
\item (ii) single $\uparrow$ ($\downarrow$) atom tunnels to occupied site at its right (left): $\Delta E_{2}\!=\!\Delta\!+\!U$; 
\item (iii) the same event as (ii) but the hopping is to the left (right): $\Delta E_{3}\!=\! U\! -\!\Delta$; 
\item  (iv) doublon hopping: $\Delta E_{4}\!=\!\Delta$
\end{itemize}
Hence, now the same three color-modulation scheme allows to realize opposite phases for the hopping of $\uparrow$ and $\downarrow$ particles of the AHM. 
\begin{align}
H^{\rm SD}_{\rm eff} \!=\! -\frac{\delta J}{2} \sum_{ \substack{j=0\cdots L\\ \sigma=0,1}} c_{j,\sigma}^\dagger e^{i (-1)^\sigma \phi |n_{\j+1,\bar\sigma}\!-\! n_{j,\bar\sigma}|} c_{i+1,\sigma} + {\rm H.c.} \,.
\label{eq:H_eff_SD}
\end{align}
Interestingly, the phase $\phi$, contrary to Model~\eqref{eq:H_eff}, has no effect on the spectrum of the model and can be gauged out in an OBC system by a simple redefinition of the fermion operators. This, however, depends on the boundary conditions and is no longer possible if we couple the two components by means of a resonant laser or microwave field. 
If we assume the particles to be trapped in a steep box-shaped trap, such that the boundaries of the systems are well defined, we may couple the boundaries~\cite{Boada2015} through spin-flips terms:
\begin{align}
H^{\rm SD}_{\rm eff} + \left ( c_{0,1}^\dagger c_{0,0} + c_{L,1}^\dagger c_{L,0} + {\rm H.c.} \right ) \,.
\end{align}

For low densities we may now interpret the system as an anyon model with single-component particles in $2L$ sites and with PBC. Most importantly, two effective single-component particles pick up a phase 
$\phi$ when exchanging their position, i.e. by traveling once around the ring. In the low density limit (i.e. if we again may neglect process (iv)) we obtain a model of, now spin-less,  
hardcore anyons on a synthetic ring:
\begin{align}
H_{\rm AHM}=\sum_{i=0\cdots 2L} \alpha_i^\dagger \alpha_{i+1} + \alpha_L^\dagger \alpha_0 + {\rm H.c.}.
\label{eq:HAHM}
\end{align}
The anyons $\alpha_i$ obey the hardcore constraint $(\alpha_i^\dagger)^2\equiv 0$, and 
the deformed exchange relation, 
\begin{align}
\alpha_j \alpha_k^\dagger + \rm{e}^{-\ii 2\phi\, {\rm sgn}(j-k)} \alpha_k^\dagger \alpha_j = \delta_{jk} \\
\alpha_j \alpha_k + \rm{e}^{-\ii 2\phi\, {\rm sgn}(j-k)} \alpha_k \alpha_j = 0
\end{align}

It is important to note that, without further interactions, Model~\eqref{eq:HAHM} is integrable. For OBC a Jordan Wigner transformation maps it to the case of free fermions and the spectrum, as well as those properties that depend on the density, are unaffected by the phase $\phi$~(see e.g. Ref.~\cite{Hao2008} and references therein for a detailed discussion on the Jordan-Wigner transformation in OBC and PBC). The quasi-momentum distribution and the single particle density matrix certainly exhibit a strong dependence on the statistics. However, an experiment will only measure the fermionic momentum distribution (since only local hoppings in the model are affected). This changes, however, for PBC. Certainly, the model is still integrable, but a mapping to free fermions leads  to a density-dependent boundary term
\begin{align}
H_{\rm AHM} = \sum_i c_i^{\dagger} c_{i+1} + \e^{\ii \phi \sum_{0<j<L-1} n _j} c_L^\dagger c_0 + {\rm H.c.}
\end{align}
We will show in the following how this effective boundary term will affect the real space density during the time evolution after a quantum quench.

\begin{figure}[b]
\begin{center}
\includegraphics[width=1\linewidth]{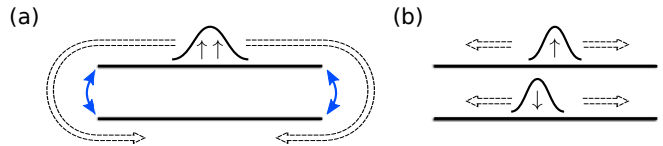}
\caption{Scheme of the experimental protocols discussed in the paper. (a) Interferometer scheme for the PBC anyon model of Sec.~\ref{sec:interferometer}. (b) Doublon expansion for the 2AHM discussed in Sec.~\ref{sec:2AHMexpand}}
\label{fig:expansion_scheme}
\end{center}
\end{figure}

\section{Dynamical probing of the exchange statistics}
\label{sec:interferometer}

The experimental setup described above, allows for the engineering of 1D anyons with an arbitrary statistical angle $0\leq \phi\leq \pi/2$. In the following, we propose an interferometer scheme 
that reveals the anyonic character by means of an expansion experiment in a small lattice system. The general idea is sketched in Fig.~\ref{fig:expansion_scheme}~(a). Initially, a spin polarized cloud of two or more particles is prepared in the center of the lattice. For concreteness we first consider exactly two (spin $\uparrow$) particles tightly confined to the two adjacent central sites. After that, we discuss the case of a larger cloud with fixed average particle number.

\begin{figure}[h]
\begin{center}
\includegraphics[width=1\columnwidth]{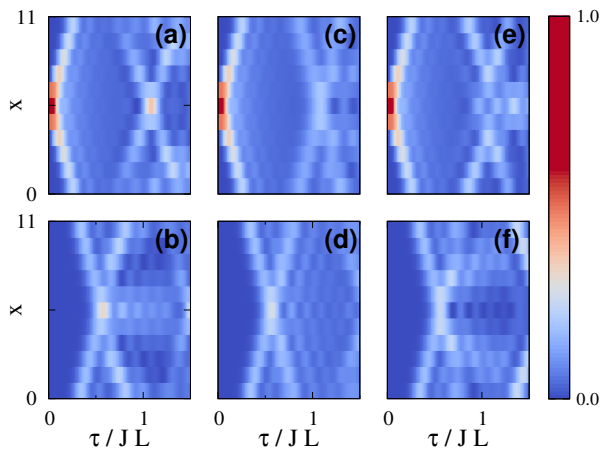}
\caption{Time evolution for (a) and (b) fermions ($\phi=0$), (c) and (d) $\phi=\pi/2$ anyons and (e) and (f) hardcore bosons ($\phi=\pi/2$). Panels (a),(c) and (e) show the $\uparrow$ component (i.e. sites $0\dots 5$ of the PBC ring system) and panels (b), (d) and (f) the $\downarrow$ component (sites $6 \dots 11$).}
\label{fig:2dexpandfree}
\end{center}
\end{figure}

\begin{figure}[h]
\begin{center}
\includegraphics[width=1\columnwidth]{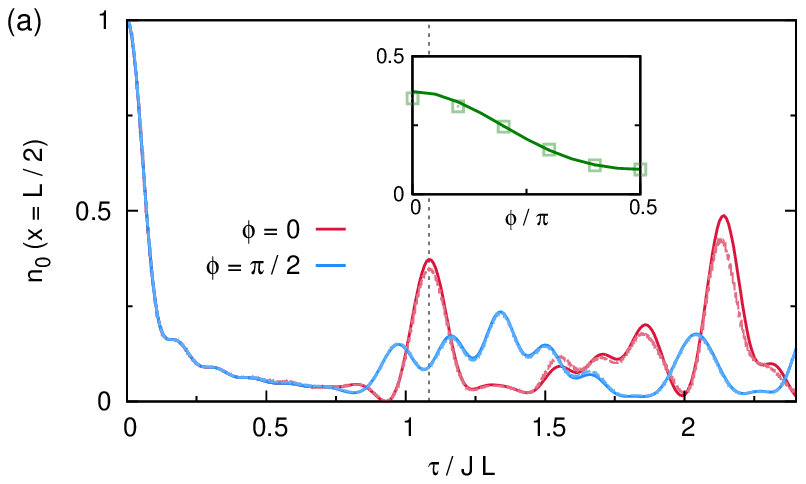}
\includegraphics[width=1\columnwidth]{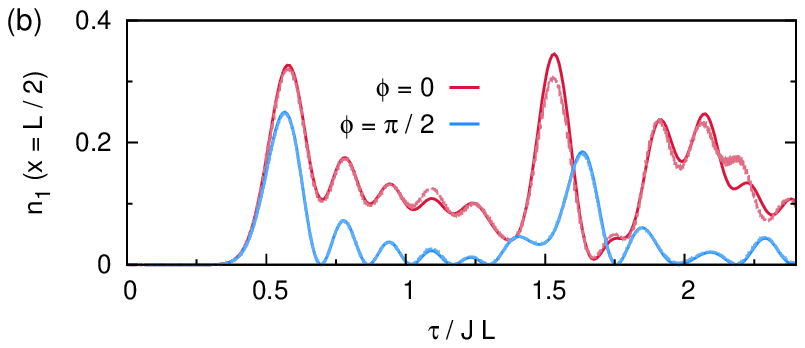}
\caption{Time evolution of the central density for the (a) spin $\uparrow$ component $n_0$ and the (b) spin $\downarrow$ component $n_1$ for the effective hardcore anyon model for fermions ($\phi=0$) and hardcore bosons ($\phi=\pi/2$). The dashed lines depict a comparison with the full three-color modulation model with $\Delta/J_0=40$, $U/J_0 = 20$~(real space length $L=12$ sites, 2 particles, $\delta J=0.5J$). The inset of (a) shows the density $n_0(x=L/2)$ for $\tau/J=L+1$ of the effective model (solid line) and the full three-color modulation simulation~(symbols) as a function of the statistical angle $\phi$.}
\label{fig:full_eff}
\end{center}
\end{figure}

\subsection{Two particle interference}

In  Fig.~\ref{fig:2dexpandfree} we show the evolution of the density of two particles for different statistical angles $\phi=0$, $\pi/4$ and $\pi/2$. The upper panels of Fig.~\ref{fig:2dexpandfree} show the evolution of the density of the spin-$\uparrow$ component $n_0$ and in the lower panels the spin-$\downarrow$ component $n_1$. 
Although the particle expansion  is diffusive and we cannot monitor the position of single particles, one may observe the emergence of an interference pattern at the center of the system after particles have in average traveled once through the whole lattice at $\tau \sim J L$. 

Fig.~\ref{fig:full_eff} depicts in detail the time evolution of the central density for both components for $\phi=0$ (effective fermions) and $\phi=\pi/2$ (effective hardcore bosons). 
For $\tau / J\gtrsim L/2$ the curves noticeably depend on the statistical angle. In particular close to the classical point of return $\tau \sim J L$ the density $n_0(L/2)$ shows a strong dependence on the statistical angle. The inset of Fig.~\ref{fig:full_eff}~(a) depicts this central spin $\uparrow$ density $n_0$ for $\tau/J=L+1$, where we observe a distinct peak for fermions and a local minimum for the bosonic case (dashed line in Fig.~\ref{fig:full_eff}). 

Fig.~\ref{fig:full_eff} also compares the evolution of the full three-color modulation Model~\eqref{eq:H_full_J(t)} and the effective PBC AHM~\eqref{eq:HAHM}. Due to higher order terms the two corresponding curves separate during the time evolution, however, for the given parameters the time evolution of Model~\eqref{eq:H_full_J(t)} recovers very well the hardcore anyon model over the full range of $\tau/J\lesssim 2L$ shown in Fig.~\ref{fig:full_eff}.

While Model~\eqref{eq:HAHM} is integrable as discussed above, for the real time evolution of the interacting two component three-color modulated Fermi-Hubbard model we employ exact diagonalization techniques in combination with a higher order Runge-Kutta method. 
	
\subsection{Fixed average particle density}

Experiments with single site resolution~\cite{bakr2009,sherson2010, endres2011, cheneau2012} may allow for the controlled initial preparation of a two particle state and the subsequent observation of the time evolution of the (possibly spin-resolved) density~\cite{Tai2016} corresponding to Fig.~\ref{fig:full_eff}. In the following we relax these conditions and analyze the possibility of an interferometrical measurement with a larger cloud with fixed average density. Initially we again assume a fully polarized sample with all particles prepared in a tight trap and ensemble-average over several realizations of the setting with fluctuating total particle densities with average density $n_{\rm avg}$ (for concreteness we choose an ensemble with $\rho(n)\sim \e^{-(n-n_{\rm avg})^2}$))

\begin{figure}[h]
\begin{center}
\includegraphics[width=1\columnwidth]{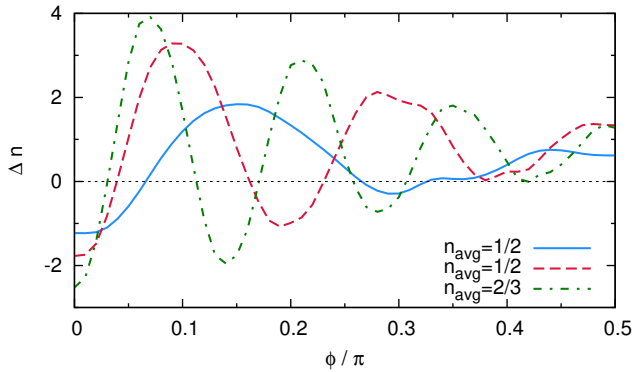}
\caption{Spin polarization as a function of the statistical angle $\phi$ for various average fillings $n_{\rm avg}$.}
\label{fig:sf_time_evolv}
\end{center}
\end{figure}

A measurement of the total spin-polarization,
\begin{align}
\Delta n = \sum_x \langle n_{0}(x) - n_{1}(x)\rangle \,,
\end{align}
may be used as a indicator of the anyonic exchange statistics. As the particles travel to the other half of the chain, they start to interfere and differences in the average populations of the two components may be measured. After long enough waiting time this difference may be quite pronounced.
In Fig.~\ref{fig:sf_time_evolv} we show the ensemble averaged value of $\Delta n$ after a fixed time $\tau / J=2L$ as a function of the statistical angle $\phi$ for different values of $n_{\rm avg}$. The curves are not a monotonous function of the phase $\phi$ and depend on $n_{\rm avg}$, however exhibit a strong dependence on the statistics of the particles.

\section{Dynamical Probing of pairing in the 2AHM}
\label{sec:2AHMexpand}

We now return to the 2AHM~\eqref{eq:2AHM} introduced in Ref.~\cite{Cardarelli2016}. We discuss how an expansion experiment may reveal the unconventional pairing properties of the 2AHM in the (pseudo) boson limit.
For the case of a pseudo-anyon Hubbard model (single component, softcore anyons) similar ideas have been discussed in Ref.~\cite{Wang2014}.

\subsection{Bound pairs in the 2AHM}

Contrary to the hardcore AHM the phase diagram of Model~\eqref{eq:2AHM} depends strongly  on the statistical angle $\phi$. Indeed as a function of $\phi$ and the filling, a plethora of ground-state phases may be found. This includes the emergence of the PP phase and a paired (singlet superconducting, SS)-phase even for vanishing interactions $\tilde{U}=0$ (see Ref.~\cite{Cardarelli2016}). A detailed analysis of the full ground state phase diagram of model~\eqref{eq:2AHM} as a function of the phase $\phi$ will be published elsewhere.

\begin{figure}[tb]
\centering
\includegraphics[width=1.\linewidth]{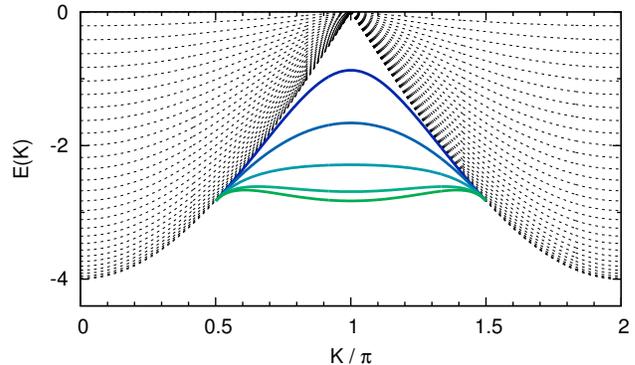}
\caption{Unconventional bound states in the two-particle spectrum of model~\eqref{eq:2AHM} (vanishing on-site interactions $\tilde{U}=0$) as a function of the total momentum of two particles $K$. Dashed lines depict the two-particle scattering continuum. Solid lines show the bound states for (from top to bottom) $\phi/\pi=0.1$, $0.2$, $0.3$, $0.4$ and $0.5$.}
\label{fig:2AHM_dilute}
\end{figure}

Both PP and SS phases can be understood from the unconventional emergence of paired states in the spectrum of the model. Following the analysis of Ref.~\cite{Cardarelli2016} and similar calculations for the softcore AHM~\cite{Zhang2017}, one observes that for a finite $\phi>0$ bound states may form in the two-particle spectrum even for vanishing on-site interactions $\tilde{U}=0$. For the 2AHM with vanishing on-site interaction term $\tilde{U}=0$, we find two-body bound states with a dispersion relation
\begin{align}
E_K = \pm 2 \sqrt{2} t  \frac{\cos(K) \cos(2\phi)+1}{\sqrt{\cos(K) (2 \cos(2\phi)-1)+1}}
\end{align}
Here $K$ is the total momentum of the two-particle solution and $\pi/2<K<3\pi/3$. For $\phi>\pi/3$ the bound state spectrum $E_K$ has a local minimum at $\pi$.  As discussed in Refs.~\cite{Cardarelli2016} due to  a quasi-condensation of bound pairs in this point PP and SS phases can form as the fractional statistics also induces an effective interaction between the anyons. Several examples of the two-particle spectrum are shown in Fig.~\ref{fig:2AHM_dilute}.

\subsection{Expansion dynamics}

\begin{figure}[t]
\centering
\includegraphics[width=1.\linewidth]{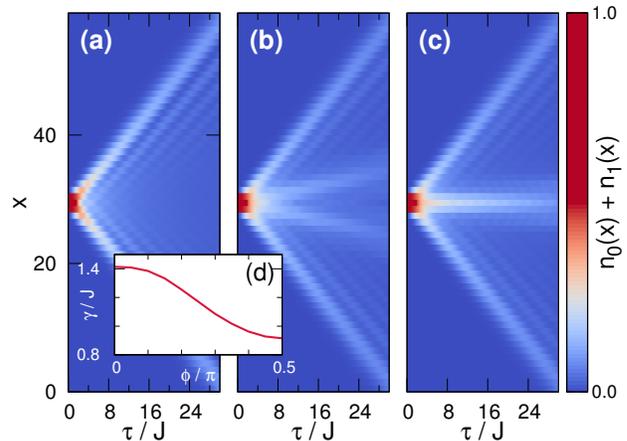}
\caption{Expansion dynamics of the total density $n_0+n_1(x)$ of the 2AHM with (a) $\phi=0$, (b) $\phi = 0.4\pi$ and (c) $\phi=\pi/2$ ($\tilde{U}=0$, $L=60$ sites) initially prepared as fully localized state (2 particles) on the two adjacent central sites (compare Fig.~\ref{fig:expansion_scheme}~(b)). Panel (d) depicts the calculated expansion rate $\gamma/J$ as a function of the statistical angle $\phi$ (see text).}
\label{fig:2AHM_texpansion}
\end{figure}

\begin{figure}[t]
\centering
\includegraphics[width=1.\linewidth]{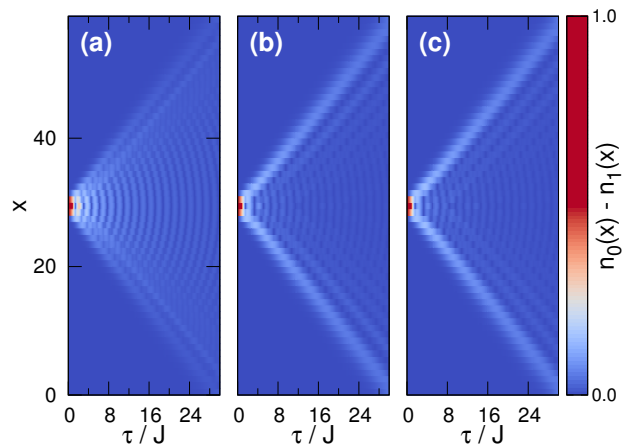}
\caption{Expansion dynamics of the spin-density $|n_0-n_1|(x)$ of the 2AHM. The parameters are the same as in Fig.~\ref{fig:2AHM_texpansion}.}
\label{fig:2AHM_texpansion_s}
\end{figure}

The formation of unconventional bound states resulting from the anyonic exchange statistics may be revealed by the characteristic expansion of a cloud of particles (now with balanced spin and OBC) into an empty lattice (Fig.~\ref{fig:expansion_scheme}~(b)).  We consider the particles initially with opposite spin on two adjacent sites in the center of an empty lattice.
In Fig.~\ref{fig:2AHM_texpansion} we show the time evolution of the real space density $n_0(x)+n_1(x)$, and in Fig.~\ref{fig:2AHM_texpansion_s} the spin $n_0(x)-n_1(x)$ for several values of $\phi$. All examples show a light-cone like ballistic expansion of the density with constant velocity independent of the statistical angle $\phi$, corresponding to single unbound particles moving  into the empty lattice. Contrary to the case of softcore anyons~\cite{Wang2014}, the light cone is symmetric for all $\phi$.

As soon as bound states can be found in the two-particle spectrum for a finite $\phi>0$ we observe a second light cone, most evident in Fig.~\ref{fig:2AHM_texpansion}~(b). As this feature is absent in the spin-density picture~(see Fig.~\ref{fig:2AHM_texpansion_s}) we conclude that it corresponds to bound pairs of particles. 
The pairs exhibit a larger effective mass due to the flatness of the bound state band~(see Fig.~\ref{fig:2AHM_dilute}) and hence the second inner light cone is much steeper. Interestingly, for our choice of initial state, the expansion of the bound-state fraction almost stops for $\phi=\pi/2$~(Fig.~\ref{fig:2AHM_dilute}~(c)).

To further quantify this expansion dynamics we monitor the evolution of the average expansion of the cloud,
\begin{align}
\Delta j(\tau) = \sqrt{\langle n_j (j-L/2)^2 \rangle (\tau)} \,, 
\end{align}
which after some initial time becomes of the form  $\Delta j (\tau)  \sim \gamma \tau$.
This expansion rate $\gamma$ is shown in Fig.~\ref{fig:2AHM_texpansion}~(d) and depends monotonously on the statistical angle $\phi$.  As expected, for free fermions ($\phi=0$, $U=0$) we find $\gamma=\sqrt{2}$. 
For finite statistical angles the expansion rate is reduced due to the enhanced tendency to form bound pairs.

\section{Conclusion and Outlook} 
\label{sec:outlook} 

In summary we have proposed a versatile experimental scheme to engineer different types of anyon Hubbard models extending the work of Ref.~\cite{Cardarelli2016}. By means of fast periodic drivings (lattice shaking or lattice depth modulation) of a two component Fermi gas in a tilted lattice a 2AHM may be engineered, whose spectrum exhibits a non trivial dependence on the statistical phase. Expansion experiments, employed for the 2AHM may reveal properties of the unconventional quantum phases of the model. In particular a clear tendency of forming bound pairs may be observed in the pseudo-boson limit, revealing the underlying mechanism of the formation of the PP-phase.

For a spin-dependent tilting the same scheme realizes a model, in which for OBC the effect of the phase may be gauged out and, hence, has no influence on the dynamics or statics of the model if one focuses on observables such as local densities. The situation  changes  drastically if one allows for Raman-assisted spin flips at the system boundaries. This scenario, may be mapped to a single component hardcore anyon model in a synthetic ring.
We have shown how fractional quantum statistics may be monitored by means of a simple interferometer scheme. The density of a cloud of expanding particles and the total spin polarization may be used to reveal clearly
the exchange statistics.

The implementation of PBC in cold atom scenarios itself has attracted considerable interest, since PBC allow for example for the observation of intriguing topological phenomena such as the Aharonv-Bohm effect or the study of persistent currents~\cite{Buttiker1983}. Experimentally ring-shaped traps~\cite{Sauer2001,Gupta2005,Ryu2007,Lesanovsky2007} have been realized and recently the implementation of PBC and further complex geometries using synthetic dimensions~\cite{Celi2014,Boada2015} or Laguerre-Gauss beams \cite{Lacki2016} has been proposed.
In this context, further interesting experimental possibilities of our proposal for the realization of PBC could include an additional phase factor $\phi_1$ to the modulations, which would allow to create a ring model penetrated by a finite flux. One may now employ this setting to analyze properties of persistent currents as function of for example interactions, particle statistics and temperature and study variations of the Drude weight. 

It is important to note that our proposal is not limited to 1D lattices, although only for this case the interpretation in terms of an anyon-model is valid. By adding an extended lattice in a second real space direction one may create a 2D or cylinder like system with unconventional correlated hoppings and fluxes.\\


\begin{acknowledgments}
We thank Marco Roncaglia and Andr\'e Eckardt for useful discussions.
We acknowledge support of the German Research Foundation DFG (projects RTG 1729 and no. SA 1031/10-1). SG also acknowledges support by the Swiss National Science Foundation under Division II.
Simulations were carried out on the cluster system at the Leibniz University of Hannover, Germany.
\end{acknowledgments}



\bibliography{references}

\end{document}